\documentclass[runningheads]{llncs}
\pdfoutput=1
\usepackage{graphicx}

\usepackage{tikz}
\usepackage{comment}
\usepackage{subcaption}
\usepackage{amsmath,amssymb} 
\usepackage{color}

\begin{document}
\pagestyle{headings}
\mainmatter

\title{Super-Resolution of Real-World Faces} 


\titlerunning{Super-Resolution of Real-World Faces}
%
\author{Saurabh Goswami\inst{1} \and
Aakanksha\inst{1} \and
Rajagopalan A. N.\inst{1}}
\authorrunning{Saurabh et al.}
%
\institute{Indian Institute of Technology, Madras, India\\
\email{\{ee18s003,ee18d405\}@smail.iitm.ac.in, raju@ee.iitm.ac.in}}

\maketitle

\begin{abstract}
Real low-resolution (LR) face images contain degradations which are too varied and complex to be captured by known downsampling kernels and signal-independent noises. So, in order to successfully super-resolve real faces, a method needs to be robust to a wide range of noise, blur, compression artifacts etc. Some of the recent works attempt to model these degradations from a dataset of real images using a Generative Adversarial Network (GAN). They generate synthetically degraded LR images and use them with corresponding real high-resolution(HR) image to train a super-resolution (SR) network using a combination of a pixel-wise loss and an adversarial loss. In this paper, we propose a two module super-resolution network where the feature extractor module extracts robust features from the LR image, and the SR module generates an HR estimate using only these robust features. We train a degradation GAN to convert bicubically downsampled clean images to real degraded images, and interpolate between the obtained degraded LR image and its clean LR counterpart. This interpolated LR image is then used along with it's corresponding HR counterpart to train the super-resolution network from end to end. Entropy Regularized Wasserstein Divergence is used to force the encoded features learnt from the clean and degraded images to closely resemble those extracted from the interpolated image to ensure robustness.
\end{abstract}

\section{Introduction}
\label{sec:intro}
Face Super-Resolution (SR) is an important preprocessing step for high-level vision tasks like facial detection and recognition. Robustness to real degradations like noise, blur, compression artifacts, etc. is one of the key aspects of the human visual system and hence highly desirable in machine vision applications as well. Even though motion blur has been useful in detecting splicing forgery in images \cite{harnessingMotionBlur}, recovering the latent motion \cite{bringingAlive}, and defocus blur has been used to infer depth from a single image \cite{depthFromMotion}, in cases where the image is very small and contains  only a single class, blur greatly reduces the recognizability of the image. Incorporating this robustness in the Super-Resolution stage itself would ease all the downstream tasks.  Unfortunately, most of the face SR methods are trained with a fixed degradation model (downsampling with a known kernel and adding noise) that is unable to capture the complexity and diversity of real degradations and hence performs poorly when applied on real degraded face images. This problem becomes more pronounced when the image is extremely small. Since most of the useful information is degraded, it further increases the ambiguity in reconstruction process. Previous methods such as \cite{supple}, \cite{superfan}, \cite{heatmap} use facial heatmaps and facial landmarks as priors to reduce ambiguity. \cite{aaaiface}, \cite{transformative} leverage autoencoders to build networks which are robust to synthetic noise and \cite{wavelet} leverage wavelet transform to train a network which is robust to gaussian noise. However, none of the above methods have been proven to be robust to real degradation except \cite{superfan}. In \cite{bulat}, a Generative Adversarial Network (GAN) was trained to generate realistically degraded Low-Resolution (LR) versions of clean High-Resolution (HR) face images and another GAN was trained to super-resolve the synthetic degraded images to their corresponding clean HR counterparts. To the best of our knowledge, this is the only previous work which super-resolves real degraded faces without the aid of any facial priors. However, we observed that \cite{bulat} produces visually different outputs for different degradations. This can be attributed to the fact that the network sees every degraded image independently and there is no explicit constraint to extract the same features from different degraded versions of the same image.\\
\indent In this paper, we focus on incorporating robustness to degradations in the task of tiny face super-resolution without the need of a face specific prior and without a dataset of \textit{degraded LR-clean HR} image pairs. Premised upon the observation that humans are remarkably adept at registering different degrdaded versions of the same image as visually similar images,  we prepend a smooth feature extractor module to our Super-Resolution (SR) module. Since our feature extractor is smooth with respect to real degradations, its output does not vary wildly when we move from clean images to degraded images. The SR module which produces clean HR images from features extracted by the smooth feature extractor, thus, produce similar images regardless of the degradation. Features which remain smooth under degradations are also features that are common between clean and degraded LR. So, our network, in essence, learns to look at features which are similar between clean and degraded LR.\\
\indent Following \cite{bulat}, we train a GAN to convert clean LR images to corresponding degraded LR images. One training iteration of our network involves two backpropagations. During the first backpropagation, we update parameters of both modules of our network to learn a super-resolution mapping from an interpolated LR (by combining clean and degraded LR) to its corresponding clean HR. The interpolation is carried out to avoid having the network overfit one of two LR domains (clean and degraded). During the second backpropagation, we minimize the Entropy Regularized Wasserstein Distance between features extracted from clean as well as degraded LR and those extracted from interpolated LR. The interpolation also helps in ensuring smoothness of the feature extractor. \\
\indent During test time, we put an image (clean or degraded) through the feature extractor module first and then feed the extracted features to the SR Module to get the corresponding super-resolved image. Since the extracted features do not change significantly between clean and degraded images, the super-resolution output for a degraded image does not change significantly from that of a clean image. We perform tests to visualise the robustness of our network as well as smoothness of the features extracted by our feature extractor.\\ 

\section{Related Works}
Single Image Super-Resolution (SISR) is a highly ill-posed inverse problem. Traditional methods mostly impose handcrafted constraints as priors to restrict the space of solutions. Early works \cite{rangeMap,resolutionEnhancement,robustComputationally} of Super-Resolution used multiple shifted low-resolution images of the same scene to retrieve the latent high resolution image conditioned on the motion between the LR images \cite{rangeMap}. The performance of these algorithms, however, are highly dependent on the motion estimates. To address this, in \cite{motionFree}, a motion free super-resolution was attempted by analytically deriving the relation for the reconstruction of the super-resolved image from its blurred and downsampled versions. With the availability of Large-scale Image Datasets and the consistent success of Convolutional Neural Networks (CNNs), learning (rather than handcrafting) a prior from a set of natural images became a possibility. Many such approaches have been explored subsequently. 
\subsection{Deep Single Image Super-Resolution}
We classify all the deep Single Image Super-Resolution (SISR) methods in two broad categories - (i) deep Paired SISR and (ii) deep Unpaired SISR. In paired SISR, corresponding pairs of LR and HR images are available and the network is evaluated on its ability to estimate an HR image given its LR counterpart. Most of the available deep paired SISR networks are trained under a setting where LR images are generated by downsampling HR images (from datasets such as Set5, Set14, DIV2K \cite{div2k}, BSD100 \cite{bsd100} etc) using a known kernel (often bicubic). These networks are trained using either a pixel wise Mean Squared Error (MSE) loss e.g. \cite{srcnn}, \cite{vdsr}, \cite{edsr}, $L_1$ loss e.g. \cite{rcan}, Charbonnier loss e.g. \cite{lapsrn} or a combination of pixel-wise $L_1$ or $L_2$ loss, perceptual loss \cite{perceptual} and adversarial loss \cite{gan} e.g. \cite{srgan}, \cite{esrgan}, \cite{fsrgan}, \cite{perceptionDistortion}. Even though these networks perform really well in terms of PSNR and SSIM, and the GAN based ones produce images that are highly realistic, these networks often fail when they are applied on real images with unseen degradations such as realistic noise and blur. Image deblurring and denoising are challenging inverse problems in their own right and most works in these areas deal with uniform blur and fixed noise models. In a recent work \cite{regionAdaptive}, the authors use deformable convolution layers to adaptively change the size of receptive field to tackle non-uniform blur. However, this work deals with only clean images. To address this, RealSR \cite{realsr} dataset was introduced in NTIRE 2019 Challenge \cite{ntire2019} containing images taken at two different focal lengths of a camera. Networks like \cite{denseres}, \cite{real1}, \cite{overfit} were trained on this dataset and are therefore robust to real degradations.\\
\indent On the other hand, in unpaired SISR, only the LR images are available in the dataset. In \cite{cincgan}, a CycleGAN \cite{cyclegan} was trained to denoise the input image and another one to finetune a pretrained super-resolution network. In \cite{timofte}, a CycleGAN was trained to generate degraded versions of clean images and a super-resolution network was then trained using pairs of synthetically degraded LR and clean HR images.\\ However, all these networks are meant for natural scenes and not faces in particular. Humans are highly sensitive to even the subtlest changes when it comes to human faces, making the task of perceptually super-resolving human faces a challenging and interesting one. 

\subsection{Deep Face SISR}
General SR networks as the ones mentioned above, often produce undesired artifacts when applied on faces. Hence, paired face SR networks often rely on face-specific prior information to subdue the artifacts and make the network focus on important features.\\
Networks like \cite{fsrgan}, \cite{superfan}, \cite{heatmap}, \cite{supple} rely on facial landmarks and heatmaps to impose additional constraints on the output whereas \cite{exemplar} leverage HR exemplars to produce high-quality HR outputs. On the other hand, networks like \cite{wavelet}, \cite{rbpnet} rely on pairs of LR and HR face images to perceptually super-resolve faces. Even though the above methods are somewhat robust to noise and occlusion, they are not equipped well enough to handle noises and blur which are as complex and as diverse as those in real images. \cite{unsupervisedClass} come close by attempting the problem of unsupervised class-specific deblurring but the blur under consideration was uniform and it still leaves the problem of denoising and super-resolution unaddressed.  \cite{transformative}, \cite{aaaiface} leverage capsule networks and transformative autoencoders to class-specifically super-resolve noisy faces but the noises are synthetic. As of yet, there seems to be no dataset with paired examples of degraded LR and clean HR images of faces available. As a result, in recent years, there has been a shift in face SISR methods from paired to unpaired. Recently, with the release of  \textit{Widerface} \cite{widerface} dataset of real low-resolution faces and the wide availability of high resolution face recognition datasets such \textit{AFLW}\cite{aflw}, \textit{VGGFace2}\cite{vggface2} and \textit{CelebAMask-HQ} \cite{CelebAMask-HQ}, Bulat et al. \cite{bulat} propose a training strategy where a High-to-Low GAN is trained to convert instances from clean HR face images to corresponding degraded LR images and a Low-to-High GAN is then trained using synthetically degraded LR images and their clean HR counterparts. This method is highly effective since it does not require facial landmarks or heatmaps for faces (as they are not available for real face images captured in the wild). \\ 
\indent However, despite producing sharp outputs, it is not very robust as different outputs are obtained for different degradations in the LR images. In order to explicitly impose robustness, we introduce a smooth feature extractor module to extract similar features from a degraded LR image and its clean LR counterpart. This enabled us to get features that are more representative of the actual face in the image and is significantly less affected by the degradations in the input.

\subsection{Robust Feature Learning}
Our work builds on the existing methods in robust feature learning.  Haoliang et al. \cite{domgen} extract robust features from multiple datasets of similar semantic contents by minimizing Maximum Mean Discrepancy (MMD) between features extracted from these datasets. Cemgil et al. \cite{smoothenc}, achieve robustness by forcing Entropy Regularized Wasserstein Distance to be low between features extracted from clean images and their noisy counterparts. None of these works handle Super-Resolution where rigorous compression using an autoencoder may hurt the reconstruction quality. We propose a method of incorporating robust feature learning in super-resolution without requiring any face specific prior information.
\section{Proposed Method}
\subsection{Motivation} Super-Resolution networks which are meant to be used on real facial images need to satisfy two criteria: (i) they need to be robust under real degradations, (ii) they should preserve the identity and pose of a face. Deep state-of-the-art super-resolution networks usually derive the LR images by bicubically downsampling HR images. Hence, an SR network trained on pairs of LR and HR images used for training fail to meet the first criterion. On the other hand, SR networks trained with real degradations fail to satisfy the second criterion. Noting the fact that the face recognition ability of us humans does not change very significantly with reasonably high degradation in images, it should be possible to find features that remain invariant under significant degradation and train a super-resolution network that would rely only on these features. Now, features which are robust to degradations would also be smooth under the said degradations. So, by enforcing explicit smoothness constraints on the extracted features, we can ensure robustness.
\subsection{Overall pipeline}
\begin{figure}[t]
    \centering
    \includegraphics[width=\textwidth]{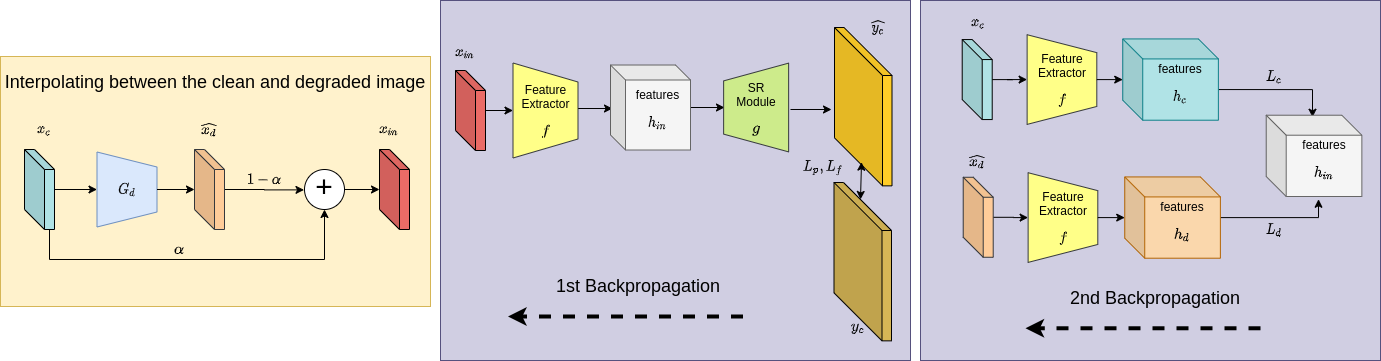}
    \caption{The proposed approach.}
    \label{fig:overall}
\end{figure}

We have a clean High-Resolution dataset $Y_c$ and a degraded Low-Resolution dataset $X_d$. We obtain clean Low-Resolution dataset, $X_c$, corresponding to $Y_c$, by downsampling every image in $Y_c$ with a bicubic downsampling kernel. So every $x_c$ in $X_c$ is a downsampled version of some $y_c$ in $Y_c$, using the equation
\begin{equation} \label{eq:lr}
x_c = (y_c \ast k)_{\downarrow s}    
\end{equation}
where, $k$ is the bicubic downsampling kernel and $s$ is the scale factor. Following \cite{bulat}, we train a Degradation GAN, $G_d$ to convert clean samples from $X_c$ to look like they have been drawn from the degraded LR dataset $X_d$. We call this synthetic degraded LR dataset $\widehat{X_d}$ and samples in this dataset $\widehat{x_d}$. So, 
\begin{equation}
    \widehat{x_d} = G_d(x_c,z) \in \widehat{X_d} \quad \forall \quad x_c \in X_c  
\end{equation}
where $z \in Z$ is an additional vector input which is sampled from a distribution $Z$ to capture the one-to-many relation between HR and degraded LR images.\\
\indent Our network basically comprises 2 modules - (i) Feature Extractor Module $(f)$ and (ii) Super-Resolution Module $(g)$. During training, we first sample an $x_c$ from $X_c$ and generate one of its degraded counterparts $\widehat{x_d} = G_d(x_c,z)$ using $G_d$. We then combine these two LR images with a mixing coefficient $\alpha$
\begin{equation} \label{eq:alpha}
    x_{in} = \alpha x_c + (1-\alpha) \widehat{x_d}
\end{equation}
where $0<\alpha<1$. We, then, put $x_{in}$ through the convolutional feature extractor $f(x)$ and the SR module $g(h)$ to estimate the corresponding clean HR output $\widehat{y_c}$ and do a backpropagation.
\begin{equation}
    h_{in} = f(x_{in}), \quad \widehat{y_c} = g(h_{in})
\end{equation}
To ensure smoothness of $f$ under real degradations, we extract features $h_c$ and $h_d$ from $x_c$ and $\widehat{x_d}$
\begin{equation}\label{eq:henc}
    h_c = f(x_c) \quad h_d = f(\widehat{x_d})
\end{equation}
and minimize the Entropy Regularized Wasserstein Distance (Sinkhorn distance) between $(h_c,h_{in})$ and $(h_d,h_{in})$ through another backpropagation. We recalculate $h_{in}$ during this operation as well. Fig.\ref{fig:overall} shows a schematic diagram of our approach.\\
\indent Here, if we use $\alpha=0$, since the entire network, during the first backpropagation, would be trained using pairs of synthetically degraded LR and clean HR samples, it may end up learning a mapping that would fail to preserve the identity of a face. However, if we take $\alpha=1$, the network may exhibit preference to the domain of clean LR images. So, we needed an input LR image which is not as sharp as $x_c$ but not as degraded as $\widehat{x_d}$ either. Since the edges in $x_c$ are much sharper than those in $\widehat{x_d}$, $x_{in}$ continues to appear reasonably clean even when $\alpha < 0.5$. This is why we do not sample $\alpha$ from a distribution since that might end up giving one domain advantage over the other and keep it fixed at $0.3$ since $\alpha=0.3$ appears to us to have struck the right balance between the two LR domains visually. \\
\indent Also, using $0<\alpha<1$, enables us to apply the smoothness constraint between $(h_c, h_{in})$ and $(h_d, h_{in})$ which is a better way to ensure smoothness than imposing smoothness constraint on pairs of $(h_c, h_d)$.

\subsection{Modeling Degradations with Degradation GAN}
Owing to the complex and diverse nature of real degradations, it is extremely difficult to mathematically model them by hand. So, following previous works \cite{bulat,timofte}, we train a GAN (termed Degradation GAN) to model real degradations. 
\subsubsection{Generator}
Our Degradation GAN Generator $G_d$, has 3 downsampling blocks, each consisting of a ResNet block followed by a $3 \times 3$ convolutional with $stride=2$, and 3 upsampling blocks each comprising ResNet blocks followed a Nearest Neighbour Upsampling layer and a $3 \times 3$ convolutional block with $stride=1$. The downsampling and upsampling paths are connected through skip connections. Our Generator takes a bicubic downsampled image $x_c$ and an $n$ dimensional random vector $z$ sampled from a normal distribution. We expand each of the $n$ dimensions of the random vector into a channel of size $H \times W$ (filled with a single value) where $H$ and $W$ are the height and width of every image. We concatenate the expanded volume with the image and feed it to the generator.

\subsubsection{Critic} We use the same discriminator used in \cite{srgan}. Since we train the degradation GAN as Wasserstein GAN \cite{wgangp}, we replace the Batch Normalization layers with Group Normalization and remove the last Sigmoid layer. Following the nomenclature, we call it critic instead of discriminator.

\subsubsection{Loss Functions}
We train the degradation GAN as a Wasserstein GAN with Gradient Penalty (WGAN-GP) \cite{wgangp}. So, the critic is trained by minimizing the following loss function:
\begin{equation}
    \resizebox{.9\hsize}{!}{$
    L_{D} = (\mathbb{E}_{x \in \widehat{X_d}}[D(x)] - \mathbb{E}_{x \in X_d}[D(x)]) + \lambda \mathbb{E}_{\widehat{x} \sim \mathbb{P}_{\widehat{x}}}[(\|\nabla_{\widehat{x}}D(x)\|_{2}-1)^2]$}
\end{equation}
where, as in \cite{wgangp}, the first term is the original critic loss and the second term is the gradient-penalty.\\
\indent To maintain the correspondence between inputs and outputs of the generator, we add a Mean Square Loss (MSE loss) term to the WGAN loss in the objective function $L_G$ of the generator.:
\begin{equation}
    L_{G} = \lambda_{WGAN} L_{WGAN} + \lambda_{MSE} L_{MSE}
\end{equation}
where,
\begin{equation}
    L_{WGAN} = -\mathbb{E}_{(x_c,z) \in (X_c,Z)}[G_d(x_c,z)]
\end{equation}
and
\begin{equation}
    L_{MSE} = \|x_c - G_d(x_c,z)\|^2
\end{equation}
\subsection{Super-Resolution using Smooth Features}
\subsubsection{Feature Extractor $f$:} 
\begin{figure}[htb]
     \centering
     \includegraphics[width=1\textwidth]{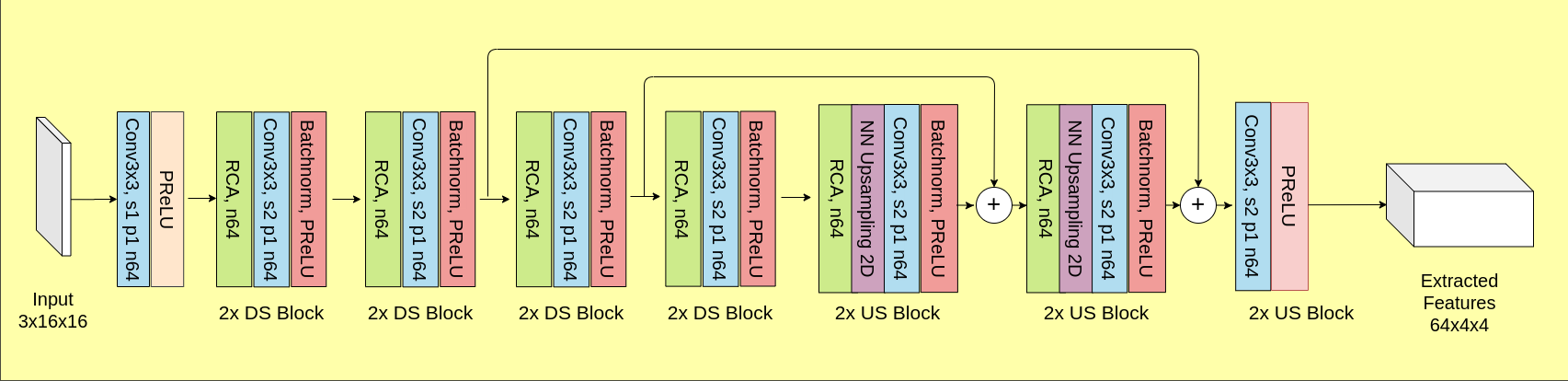}
     \caption{Feature Extractor $f$.}
     \label{fig:featex}
\end{figure}

Our feature extractor consists of $4$ Residual Channel Attention (RCA) downsampling and $2$ upsampling blocks. As shown in Fig. \ref{fig:featex}, there are 2 skip connections. It is a fully convolutional module which takes an LR image of dimension $3 \times 16 \times 16$ at the input and produces a feature volume of dimension $64 \times 4 \times 4$. In Fig. \ref{fig:featex}, `RCA, n64' denotes an RCA block with $64$ output channels and `Conv3x3, s2 p1 n64' denotes a $3 \times 3$ convolutional layer with $stride=2$, $padding=1$ and $64$ output channels.

\subsubsection{Super-Resolution Module $g$:} 
\begin{figure}
    \centering
    \includegraphics[width=\textwidth]{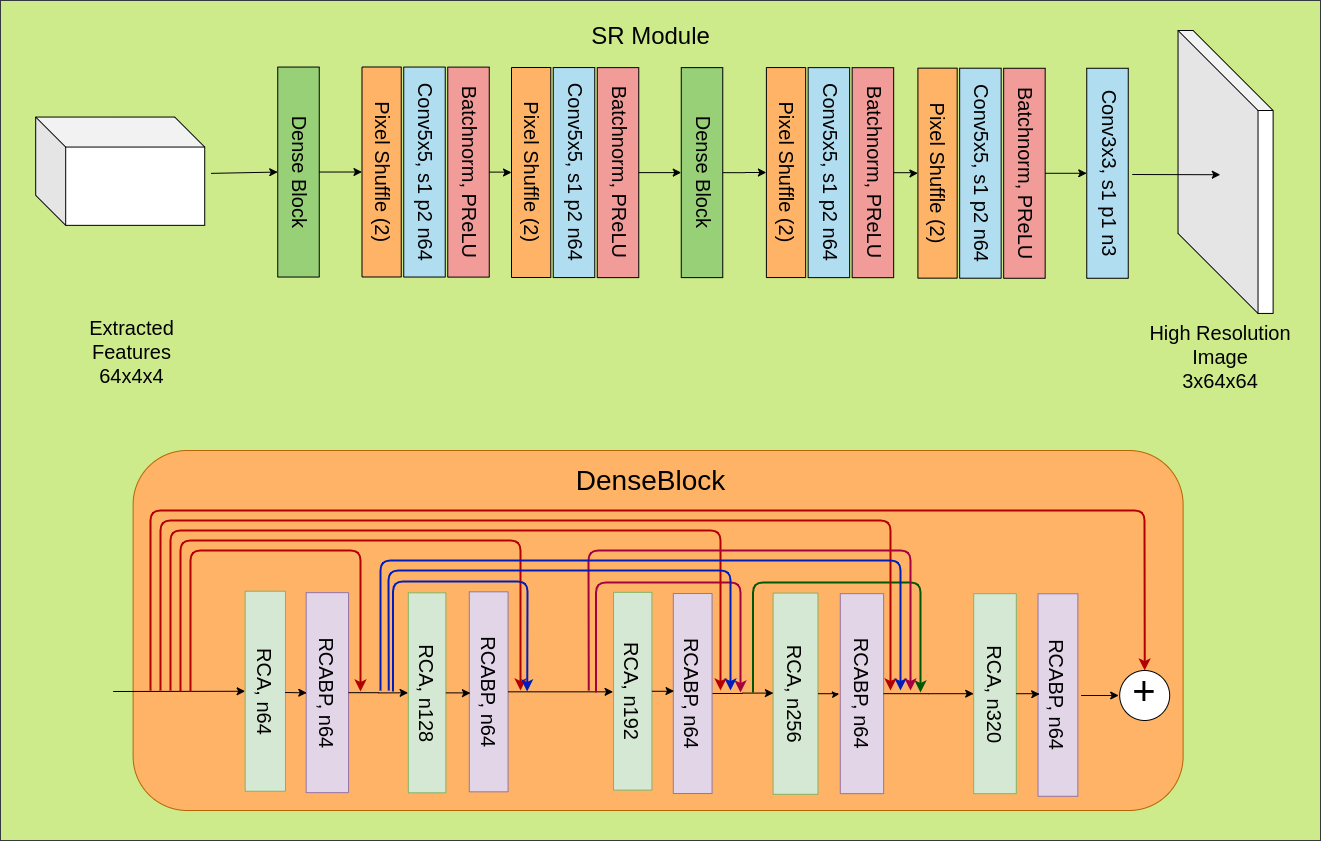}
    \caption{Architecture of SR Module $g$ and DenseBlock.}
    \label{fig:srmodule}
\end{figure}

Our Super-Resolution module consists of $6$ upsampling blocks and $2$ DenseBlocks as shown in Fig. \ref{fig:srmodule}. The upsampling blocks comprise a Pixel-Shuffle layer, a convolution layer, a Batch-Normalization layer and a PReLU layer. The DenseBlocks contain a number of Residual Channel Attention (RCA) blocks and Residual Channel Attention Back-Projection (RCABP) blocks connected in a dense fashion as in \cite{denseres}. In Fig. \ref{fig:srmodule}, `Pixel Shuffle ($2$)' denotes 2x pixel-shuffle upsampling layer and `RCABP, n64' stands for an RCABP block with $64$ heatmaps at the output.

\indent
During one forward pass, we pass a minibatch of $x_{in}$ through our feature extractor $f$ to produce the feature volume $h_{in}$. We put $h_{in}$ through our Super-Resolution module $g$ to produce a high resolution estimate $\widehat{y_c}$ and do a back propagation through both $g$ and $f$. This ensures that the features are useful for SR. Since $x_{in}$ is neither as clean as $x_c$ nor as severely degraded as $\widehat{x_d}$, the possibility of our SR network being biased to any one of the domains is eliminated.\\
\indent After the first backpropagation, we put one minibatch each of $x_c, \widehat{x_d}$ and $x_{in}$ (again) through $f$, as shown in Eq. \ref{eq:henc}, and calculate the Sinkhorn Distance \cite{sinkhorn} (which calculates the Entropy Regularized Wasserstein Divergence) between $(h_c, h_{in})$ and $(h_d, h_{in})$,
 \begin{equation}
     L_c = Sinkhorn(h_c, h_{in}), \quad L_d = Sinkhorn(h_d, h_{in})
 \end{equation}
 Using a combination of $L_c$ and $L_d$ as a loss function, we backpropagate through $f$ one more time to enforce smoothness under degradations.\\
 \indent Like our Degradation GAN, we train our robust super-resolution network (during the first back propagation) like a Wasserstein GAN. So, the objective function here is a combination of adversarial loss $(L_{adv})$, pixel-level $L_1$ loss $(L_p)$ and a perceptual loss \cite{perceptual} $(L_f)$ computed between features extracted from the estimated $(\widehat{y_c})$ and ground-truth $(y_c)$ HR images through a subset of VGG16 network. Hence, the overall objective function optimized during the first back propagation is
 \begin{equation}
     L_{sr} = \lambda_p L_p + \lambda_f L_f + \lambda_{adv} L_{adv}
 \end{equation}
 where,
 \begin{align}
     L_p &= \|y_c - \widehat{y_c}\|_1    \\
     L_f &= \|f_{vgg}(y_c) - f_{vgg}(\widehat{y_c})\|_1 \\
     L_{adv} &= -\mathbb{E}_{x_{in} \sim \widehat{\mathbb{P}_x}}[D_{sr}(g(f(x_{in})))] 
 \end{align}
 with $f_{vgg}$ being a subset of VGG16 network, $\mathbb{P}_x$ being the distribution described by $x_{in}$ and $D_{sr}$ being the critic comparing the generated HR images with the ground-truth HR images. The architecture of $D_{sr}$ is same as the critic of degradation GAN and it is trained with the following loss function:
 \begin{equation}
    \resizebox{.9\hsize}{!}{$
    L_{DSR} = (\mathbb{E}_{\widehat{y_c} \sim \mathbb{P}_{y}}[D_{sr}(\widehat{y_c})] - \mathbb{E}_{y_c \in Y_c}[D(y_c)]) + \lambda \mathbb{E}_{\widehat{y} \sim \mathbb{P}_{\widehat{y}}}[(\|\nabla_{\widehat{y}}D(\widehat{y})\|_{2}-1)^2]$}
\end{equation}
where $\mathbb{P}_y$ is the distribution generated by the outputs of our network and $\mathbb{P}_{\widehat{y}}$ is the distribution of samples interpolated between $\widehat{y_c}$ and $y_c$.  \\
\indent For the second back propagation, we optimize a combination of the Sinkhorn Distances mentioned earlier
\begin{equation}
    L_{robust} = \lambda_c L_c + \lambda_d L_d
\end{equation}
Since the second backpropagation is only through $f$, it does not directly affect the mapping learnt by $g$ and only makes $f$ smooth under degradations.

\section{Experiments}
\label{sec:exp}
\subsection{Training Details} We use two-time step update for both our Degradation GAN and Robust Super-Resolution Network. For both $D$ and $D_{sr}$, we start with a learning rate of $4 \times 10^{-4}$ and decrease them by a factor of $0.5$ after every $10000$ iterations. For all the other networks $(G_d, f, g)$ we set the initial training at $10^{-4}$ and decay it by a factor of $0.5$ after every $10000$ iterations.\\
\indent For all networks, we use Adam Optimizer with $\beta_1 = 0.0$ and $\beta_2=0.9$. For every $5$ updates of discriminators, we update the corresponding generator networks once. We try out a number of different values of $\lambda$ and the ones that worked best for us are $[\lambda_{WGAN} = 0.05 , \lambda_{MSE} = 1 , \lambda_p = 1 , \lambda_f = 0.5 , \lambda_{adv} = 0.05 , \lambda_c = 0.3, \lambda_d = 0.7]$. For $G_d$, we sample $z$ from a $16-$dimensional multivariate normal distribution with zero mean and unit standard deviation.

\subsection{Datasets}
We train our network for $4\times$ super-resolution ($s=4$). However, our robustness strategy is not scale dependent. For training our network, we used two datasets: one with degraded images and the other with clean images. To make the degraded image dataset, we randomly sample 153446 images from the \textit{Widerface} \cite{widerface} dataset. This dataset contains face images with a wide range of degradations such as: varying degrees of noise, extreme poses and expressions, occlusions, skew, non-uniform blur etc. We use 138446 of these images for training and 15000 for testing. While compiling the clean dataset, to make sure it is diverse enough in terms of poses, occlusions, skin colours and expressions, we combined the entire \textit{AFLW} \cite{aflw} dataset with 60000 images from \textit{CelebAMask-HQ} \cite{CelebAMask-HQ} dataset and 100000 images from \textit{VGGFace2} \cite{vggface2} dataset. To obtain clean LR images, we simply downsample images from the clean dataset.
 
 \subsection{Results}
    \begin{figure}[!t]
                \centering
                \begin{tabular}{ccccccc}
                    LR & \includegraphics[width=0.10\textwidth, height=0.10\textwidth]{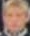} & \includegraphics[width=0.10\textwidth, height=0.10\textwidth]{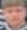} & \includegraphics[width=0.10\textwidth, height=0.10\textwidth]{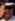} & \includegraphics[width=0.10\textwidth, height=0.10\textwidth]{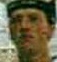} & \includegraphics[width=0.10\textwidth, height=0.10\textwidth]{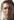} & \includegraphics[width=0.10\textwidth, height=0.10\textwidth]{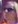}\\
                    ESRGAN & \includegraphics[width=0.10\textwidth]{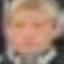} & \includegraphics[width=0.10\textwidth]{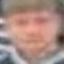} & \includegraphics[width=0.10\textwidth]{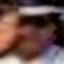} & \includegraphics[width=0.10\textwidth]{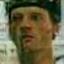} & \includegraphics[width=0.10\textwidth]{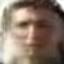} & \includegraphics[width=0.10\textwidth]{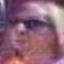}\\
                    Bulat et al. & \includegraphics[width=0.10\textwidth]{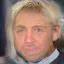} & \includegraphics[width=0.10\textwidth]{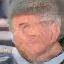} & \includegraphics[width=0.10\textwidth]{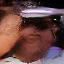} & \includegraphics[width=0.10\textwidth]{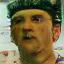} & \includegraphics[width=0.10\textwidth]{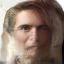} & \includegraphics[width=0.10\textwidth]{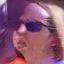}\\
                    Ours & \includegraphics[width=0.10\textwidth]{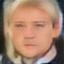} & \includegraphics[width=0.10\textwidth]{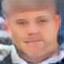} & \includegraphics[width=0.10\textwidth]{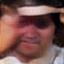} & \includegraphics[width=0.10\textwidth]{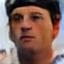} & \includegraphics[width=0.10\textwidth]{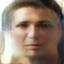} & \includegraphics[width=0.10\textwidth]{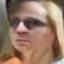}\\
                \end{tabular}
                \caption{Results for Real-Degraded Dataset.}
                \label{fig:real_deg}
        \end{figure}
     \textbf{Real-Degraded Dataset:} This dataset contains $15000$ images from the \textit{Widerface} Dataset. Performance on this dataset will dictate how effective our method is in super-resolving real degraded facial images.\\
     \indent As shown in Fig. \ref{fig:real_deg}, our method is able to super-resolve real degraded faces. The outputs of \cite{bulat} contain undesired artifacts and sometimes exhibit identity discrepency as well. ESRGAN \cite{esrgan} is able to maintain the identity but the outputs are not sharp. Since we do not have ground-truth HR images for these LR images, we can not compute PSNR/SSIM. So, we use Fretchet Inception Distance (FID) as a metric to assess how close the output is to the target distribution of sharp images. Table \ref{table:fid} shows the FIDs of \cite{esrgan}, \cite{bulat} and our method computed over $15000$ images. Lower FID denotes better adherence to target distribution and hence sharper output.
     \begin{table}[h]
         \centering
         \scriptsize
         \begin{tabular}{|c|c|}
              \hline
              Method & FID \\
              \hline
              \hline
              ESRGAN \cite{esrgan} & 139.2599\\
              \hline
              Bulat et al. \cite{bulat} & \textbf{74.2798}\\
              \hline
              Ours & 77.1359\\
              \hline
         \end{tabular}
         \caption{Comparison of FID.}
         \label{table:fid}
     \end{table}
     As shown in Table \ref{table:fid}, our method performs very close to \cite{bulat} in terms of realness of the output and at the same time, maintains a fixed output under varying degradations. So, our method is robust and at the same time, effective on real degraded faces.

\section{Conclusion}
We propose a robust super-resolution network that would give consistent output under a wide range of degradations. We train a feature extractor that is able to extract similar features from both bicubically downsampled images and their corresponding realistically degraded counterparts. We perform robustness test to put our claim of robustness to test and smoothness test to visualize the variation in extracted features as we gradually move from a clean to a degraded LR image. There is still room to improve our network for better performance in terms of PSNR/SSIM. In our future works, we will attempt to address this. Refined and complete version of this work appeared in the 'Adversarial Robustness in the Real World' in the European Conference on Computer Vision 2020.
\clearpage
%
%


\begin{thebibliography}{10}
\providecommand{\url}[1]{\texttt{#1}}
\providecommand{\urlprefix}{URL }
\providecommand{\doi}[1]{https://doi.org/#1}

\bibitem{div2k}
Agustsson, E., Timofte, R.: Ntire 2017 challenge on single image
  super-resolution: Dataset and study. In: The IEEE Conference on Computer
  Vision and Pattern Recognition (CVPR) Workshops (July 2017)

\bibitem{bsd100}
Arbelaez, P., Maire, M., Fowlkes, C., Malik, J.: Contour detection and
  hierarchical image segmentation. IEEE Trans. Pattern Anal. Mach. Intell.
  \textbf{33}(5),  898--916 (May 2011). \doi{10.1109/TPAMI.2010.161},
  \url{http://dx.doi.org/10.1109/TPAMI.2010.161}

\bibitem{rangeMap}
Bhavsar, A.V., Rajagopalan, A.N.: Range map superresolution-inpainting, and
  reconstruction from sparse data. Computer Vision and Image Understanding
  \textbf{116}(4),  572--591 (2012)

\bibitem{resolutionEnhancement}
Bhavsar, A.V., Rajagopalan, A.: Resolution enhancement in multi-image stereo.
  IEEE transactions on pattern analysis and machine intelligence
  \textbf{32}(9),  1721--1728 (2010)

\bibitem{superfan}
Bulat, A., Tzimiropoulos, G.: Super-fan: Integrated facial landmark
  localization and super-resolution of real-world low resolution faces in
  arbitrary poses with gans. CoRR  \textbf{abs/1712.02765} (2017),
  \url{http://arxiv.org/abs/1712.02765}

\bibitem{bulat}
Bulat, A., Yang, J., Tzimiropoulos, G.: To learn image super-resolution, use a
  gan to learn how to do image degradation first. In: Ferrari, V., Hebert, M.,
  Sminchisescu, C., Weiss, Y. (eds.) Computer Vision -- ECCV 2018. pp.
  187--202. Springer International Publishing, Cham (2018)

\bibitem{ntire2019}
Cai, J., Gu, S., Timofte, R., Zhang, L.: Ntire 2019 challenge on real image
  super-resolution: Methods and results. In: Proceedings of the IEEE Conference
  on Computer Vision and Pattern Recognition Workshops (2019)

\bibitem{realsr}
Cai, J., Zeng, H., Yong, H., Cao, Z., Zhang, L.: Toward real-world single image
  super-resolution: A new benchmark and a new model. In: Proceedings of the
  IEEE International Conference on Computer Vision (2019)

\bibitem{vggface2}
Cao, Q., Shen, L., Xie, W., Parkhi, O.M., Zisserman, A.: Vggface2: A dataset
  for recognising faces across pose and age. In: International Conference on
  Automatic Face and Gesture Recognition (2018)

\bibitem{smoothenc}
Cemgil, T., Ghaisas, S., Dvijotham, K.D., Kohli, P.: Adversarially robust
  representations with smooth encoders. In: International Conference on
  Learning Representations (2020),
  \url{https://openreview.net/forum?id=H1gfFaEYDS}

\bibitem{rbpnet}
Chen, X., Wang, X., Lu, Y., Li, W., Wang, Z., Huang, Z.: Rbpnet: An asymptotic
  residual back-projection network for super-resolution of very low-resolution
  face image. Neurocomputing  \textbf{376},  119 -- 127 (2020).
  \doi{https://doi.org/10.1016/j.neucom.2019.09.079},
  \url{http://www.sciencedirect.com/science/article/pii/S0925231219313530}

\bibitem{fsrgan}
Chen, Y., Tai, Y., Liu, X., Shen, C., Yang, J.: Fsrnet: End-to-end learning
  face super-resolution with facial priors. CoRR  \textbf{abs/1711.10703}
  (2017), \url{http://arxiv.org/abs/1711.10703}

\bibitem{sinkhorn}
Cuturi, M.: Sinkhorn distances: Lightspeed computation of optimal
  transportation distances (2013)

\bibitem{exemplar}
Dogan, B., Gu, S., Timofte, R.: Exemplar guided face image super-resolution
  without facial landmarks. CoRR  \textbf{abs/1906.07078} (2019),
  \url{http://arxiv.org/abs/1906.07078}

\bibitem{srcnn}
Dong, C., Loy, C.C., He, K., Tang, X.: Image super-resolution using deep
  convolutional networks. CoRR  \textbf{abs/1501.00092} (2015),
  \url{http://arxiv.org/abs/1501.00092}

\bibitem{real1}
Du, C., Zewei, H., Anshun, S., Jiangxin, Y., Yanlong, C., Yanpeng, C., Siliang,
  T., Ying~Yang, M.: Orientation-aware deep neural network for real image
  super-resolution. In: The IEEE Conference on Computer Vision and Pattern
  Recognition (CVPR) Workshops (June 2019)

\bibitem{overfit}
Feng, R., Gu, J., Qiao, Y., Dong, C.: Suppressing model overfitting for image
  super-resolution networks. In: The IEEE Conference on Computer Vision and
  Pattern Recognition (CVPR) Workshops (June 2019)

\bibitem{gan}
Goodfellow, I.J., Pouget-Abadie, J., Mirza, M., Xu, B., Warde-Farley, D.,
  Ozair, S., Courville, A., Bengio, Y.: Generative adversarial networks (2014)

\bibitem{wgangp}
Gulrajani, I., Ahmed, F., Arjovsky, M., Dumoulin, V., Courville, A.C.: Improved
  training of wasserstein gans. CoRR  \textbf{abs/1704.00028} (2017),
  \url{http://arxiv.org/abs/1704.00028}

\bibitem{wavelet}
{Huang}, H., {He}, R., {Sun}, Z., {Tan}, T.: Wavelet-srnet: A wavelet-based cnn
  for multi-scale face super resolution. In: 2017 IEEE International Conference
  on Computer Vision (ICCV). pp. 1698--1706 (2017)

\bibitem{denseres}
{Jang}, D., {Park}, R.: Densenet with deep residual channel-attention blocks
  for single image super resolution. In: 2019 IEEE/CVF Conference on Computer
  Vision and Pattern Recognition Workshops (CVPRW). pp. 1795--1803 (2019)

\bibitem{perceptual}
Johnson, J., Alahi, A., Fei-Fei, L.: Perceptual losses for real-time style
  transfer and super-resolution (2016)

\bibitem{vdsr}
Kim, J., Lee, J.K., Lee, K.M.: Accurate image super-resolution using very deep
  convolutional networks. CoRR  \textbf{abs/1511.04587} (2015),
  \url{http://arxiv.org/abs/1511.04587}

\bibitem{lapsrn}
Lai, W., Huang, J., Ahuja, N., Yang, M.: Fast and accurate image
  super-resolution with deep laplacian pyramid networks. CoRR
  \textbf{abs/1710.01992} (2017), \url{http://arxiv.org/abs/1710.01992}

\bibitem{srgan}
Ledig, C., Theis, L., Huszar, F., Caballero, J., Aitken, A.P., Tejani, A.,
  Totz, J., Wang, Z., Shi, W.: Photo-realistic single image super-resolution
  using a generative adversarial network. CoRR  \textbf{abs/1609.04802} (2016),
  \url{http://arxiv.org/abs/1609.04802}

\bibitem{CelebAMask-HQ}
Lee, C.H., Liu, Z., Wu, L., Luo, P.: Maskgan: Towards diverse and interactive
  facial image manipulation. arXiv preprint arXiv:1907.11922  (2019)

\bibitem{domgen}
Li, H., Jialin~Pan, S., Wang, S., Kot, A.C.: Domain generalization with
  adversarial feature learning. In: The IEEE Conference on Computer Vision and
  Pattern Recognition (CVPR) (June 2018)

\bibitem{edsr}
Lim, B., Son, S., Kim, H., Nah, S., Lee, K.M.: Enhanced deep residual networks
  for single image super-resolution. CoRR  \textbf{abs/1707.02921} (2017),
  \url{http://arxiv.org/abs/1707.02921}

\bibitem{timofte}
Lugmayr, A., Danelljan, M., Timofte, R.: Unsupervised learning for real-world
  super-resolution (2019)

\bibitem{aflw}
Martin~Koestinger, Paul~Wohlhart, P.M.R., Bischof, H.: {Annotated Facial
  Landmarks in the Wild: A Large-scale, Real-world Database for Facial Landmark
  Localization}. In: {Proc. First IEEE International Workshop on Benchmarking
  Facial Image Analysis Technologies} (2011)

\bibitem{unsupervisedClass}
Nimisha, T.M., Sunil, K., Rajagopalan, A.: Unsupervised class-specific
  deblurring. In: Proceedings of the European Conference on Computer Vision
  (ECCV). pp. 353--369 (2018)

\bibitem{depthFromMotion}
Paramanand, C., Rajagopalan, A.N.: Depth from motion and optical blur with an
  unscented kalman filter. IEEE Transactions on Image Processing
  \textbf{21}(5),  2798--2811 (2011)

\bibitem{regionAdaptive}
Purohit, K., Rajagopalan, A.: Region-adaptive dense network for efficient
  motion deblurring. In: Proceedings of the AAAI Conference on Artificial
  Intelligence. vol.~34, pp. 11882--11889 (2020)

\bibitem{bringingAlive}
Purohit, K., Shah, A., Rajagopalan, A.: Bringing alive blurred moments. In:
  Proceedings of the IEEE/CVF Conference on Computer Vision and Pattern
  Recognition. pp. 6830--6839 (2019)

\bibitem{motionFree}
Rajagopalan, A.N., Kiran, V.P.: Motion-free superresolution and the role of
  relative blur. JOSA A  \textbf{20}(11),  2022--2032 (2003)

\bibitem{harnessingMotionBlur}
Rao, M.P., Rajagopalan, A., Seetharaman, G.: Harnessing motion blur to unveil
  splicing. IEEE transactions on information forensics and security
  \textbf{9}(4),  583--595 (2014)

\bibitem{robustComputationally}
Suresh, K.V., Rajagopalan, A.N.: Robust and computationally efficient
  superresolution algorithm. JOSA A  \textbf{24}(4),  984--992 (2007)

\bibitem{perceptionDistortion}
Vasu, S., Thekke~Madam, N., Rajagopalan, A.: Analyzing perception-distortion
  tradeoff using enhanced perceptual super-resolution network. In: Proceedings
  of the European Conference on Computer Vision (ECCV) Workshops. pp.~0--0
  (2018)

\bibitem{esrgan}
Wang, X., Yu, K., Wu, S., Gu, J., Liu, Y., Dong, C., Qiao, Y., Loy, C.C.:
  Esrgan: Enhanced super-resolution generative adversarial networks. In:
  Leal-Taix{\'e}, L., Roth, S. (eds.) Computer Vision -- ECCV 2018 Workshops.
  pp. 63--79. Springer International Publishing, Cham (2019)

\bibitem{aaaiface}
Xin, J., Wang, N., Jiang, X., Li, J., Gao, X., Li, Z.: Facial attribute
  capsules for noise face super resolution (2020)

\bibitem{widerface}
Yang, S., Luo, P., Loy, C.C., Tang, X.: Wider face: A face detection benchmark.
  In: IEEE Conference on Computer Vision and Pattern Recognition (CVPR) (2016)

\bibitem{supple}
{Yu}, X., {Fernando}, B., {Hartley}, R., {Porikli}, F.: Super-resolving very
  low-resolution face images with supplementary attributes. In: 2018 IEEE/CVF
  Conference on Computer Vision and Pattern Recognition. pp. 908--917 (2018)

\bibitem{transformative}
{Yu}, X., {Porikli}, F.: Hallucinating very low-resolution unaligned and noisy
  face images by transformative discriminative autoencoders. In: 2017 IEEE
  Conference on Computer Vision and Pattern Recognition (CVPR). pp. 5367--5375
  (2017)

\bibitem{heatmap}
Yu, X., Fernando, B., Ghanem, B., Porikli, F., Hartley, R.: Face
  super-resolution guided by facial component heatmaps. In: Ferrari, V.,
  Hebert, M., Sminchisescu, C., Weiss, Y. (eds.) Computer Vision -- ECCV 2018.
  pp. 219--235. Springer International Publishing, Cham (2018)

\bibitem{cincgan}
{Yuan}, Y., {Liu}, S., {Zhang}, J., {Zhang}, Y., {Dong}, C., {Lin}, L.:
  Unsupervised image super-resolution using cycle-in-cycle generative
  adversarial networks. In: 2018 IEEE/CVF Conference on Computer Vision and
  Pattern Recognition Workshops (CVPRW). pp. 814--81409 (2018)

\bibitem{rcan}
Zhang, Y., Li, K., Li, K., Wang, L., Zhong, B., Fu, Y.: Image super-resolution
  using very deep residual channel attention networks. CoRR
  \textbf{abs/1807.02758} (2018), \url{http://arxiv.org/abs/1807.02758}

\bibitem{cyclegan}
Zhu, J., Park, T., Isola, P., Efros, A.A.: Unpaired image-to-image translation
  using cycle-consistent adversarial networks. CoRR  \textbf{abs/1703.10593}
  (2017), \url{http://arxiv.org/abs/1703.10593}

\end{thebibliography}
\end{document}